\documentclass[prl,
twocolumn,
secnumarabic,showpacs]{revtex4}

\usepackage{epsfig}
\usepackage{graphicx}

\def\RR{\rm \hbox{I\kern-.2em\hbox{R}}}
\def\NN{\rm \hbox{I\kern-.2em\hbox{N}}}
\def\ZZ{\rm {{\rm Z}\kern-.28em{\rm Z}}}
\def\CC{\rm \hbox{C\kern -.5em {\raise .32ex \hbox{$\scriptscriptstyle
|$}}\kern
-.22em{\raise .6ex \hbox{$\scriptscriptstyle |$}}\kern .4em}}

\def\s{{\cal S}}
\def\b{\begin{equation}}
\def\e{\end{equation}}

\usepackage{pifont}

\newcommand{\ligne}{\protect\rule[0.6ex]{2.0em}{0.1ex}}

\usepackage{amssymb}

\begin{document}   

\title{ A multifractal random walk}

\author{E. Bacry}
\affiliation{Centre de Math\'ematiques Appliqu\'ees, Ecole Polytechnique,
91128 Palaiseau Cedex, France}
\author{J. Delour}
\author{J.F. Muzy} 
\affiliation{Centre de Recherche Paul Pascal, Avenue Schweitzer 33600
Pessac, France}

\date{\today}

\begin{abstract}

We introduce a  class of multifractal processes, referred to as Multifractal
Random Walks (MRWs). To our knowledge,
it is the first multifractal processes with continuous dilation invariance
properties and stationary increments.
MRWs are very attractive alternative processes to classical cascade-like
multifractal models since
they do not involve any particular scale ratio. The MRWs are indexed by few
parameters
that are shown to control in a very direct way the
multifractal spectrum and the correlation structure of the increments. We
briefly explain how, in the same way, one
can build stationary multifractal processes or positive random measures.

\end{abstract}


\maketitle

 Multifractal models have been used to account for scale
invariance properties of various objects in very different domains ranging
from the energy dissipation or the
velocity field in turbulent flows to financial data. The scale invariance
properties of a deterministic
fractal function $f(t)$ are generally characterized by the exponents
$\zeta_q$ which govern the power law scaling of the absolute moments of its
fluctuations, i.e.,
\b
\label{moment}
m(q,l) = K_q l^{\zeta_q},
\e
where, for instance, one can choose $m(q,l) = \sum_t |f(t+l)-f(t)|^q$.
When the exponents $\zeta_q$ are linear in $q$, a single scaling exponent
$H$ is involved. One has $\zeta_q = qH$ and $f(t)$ is said to be {\em
monofractal}. If
the function $\zeta_q$ is no longer linear in $q$, $f(t)$ is said to be
{\em multifractal}.
In the case of a stochastic process $X(t)$ with stationary increments, these
definitions are naturally extended using
\b
m(q,l) = E\left(|\delta_l X(t)|^q\right) = E\left(|X(t+l)-X(t)|^q\right),
\e
where $E$ stands for the expectation.
Some very popular monofractal stochastic processes are the so-called {\em
self-similar processes} \cite{bTaqq}. They
are defined as processes $X(t)$ which have stationary increments and which
verify (in law)
\b
\label{ss}
\delta_{\lambda l} X(t) =
\lambda^H \delta_l X(t), ~~\forall l,\lambda >0.
\e
Widely used examples of such processes are fractional Brownian
motions (fBm) and Levy walks. One reason for their success is that, as it is
generally the case in experimental
time-series, they do not involve any particular scale ratio (i.e., there
is no constraint on $l$ or $\lambda$ in
Eq. (\ref{ss})). In the same spirit, one can try to build
multifractal processes which do not involve any
particular scale ratio. A common approach originally proposed
by several authors in the field of fully developed turbulence
\cite{nov,sl,FP,dg,castaing}, has been to describe such
processes in terms of differential
equations, in the scale domain, describing the cascading process that rules
how the fluctuations evolves when
going from coarse to fine scales.
One can state that the fluctuations
at scales $l$ 
and $\lambda l$ ($\lambda < 1$) are related (for fixed $t$) through the
infinitesimal ($\lambda = 1-\eta$ with $\eta << 1$) cascading rule
\b
\label{castaingd}
\delta_{\lambda l} X(t) = W_\lambda \delta_l X(t)
\e
where $W_\lambda$ is a stochastic variable which depends only on $\lambda$.
Let us note that this latter equation
can be simply seen as a generalization of Eq. (\ref{ss}) with
$H$ being stochastic.
Since Eq. (\ref{castaingd}) can be iterated, it
implicitely imposes the random variable $W_\lambda$ to have a log infinitely
divisible law. However, according to our knowledge,
nobody has succeeded in building
effectively such processes yet, mainly because of the peculiar
constraints in the time-scale half-plane.
The integral equation corresponding to this infinitesimal approach
has been proposed by
Castaing {\em et al.} \cite{castaing}. It relates the probability density
function (pdf)  $P_l(\delta X)$ of
$\delta_l X$ to the pdf $G_{\lambda l,l}$ of $\ln W_\lambda$ :
\b
\label{castaing}
P_{\lambda l}(\delta X) = \int G_{\lambda l,l}(u) e^{-u} P_{l}(e^{-u} \delta
X) du.
\e
The {\em self-similarity}  kernel
$G_{\lambda l,l}$ satisfies the same iterative rule as
$W_\lambda$ which implies that its Fourier transform is of the form ${\hat
G}_{\lambda l,l}(k) = {\hat
G}^\lambda(k)$. Thus one can easily show that the $q$ order absolute moments
at scale $l$ scales like
\b
m(q,l) = \hat G_{l,L}(-iq)m(q,L) = m(q,L) \left(\frac{l}{L}\right)^{F(-iq)},
\e
where $F =\ln \hat G$ refers to the
cumulant generating function of $\ln W$ \cite{castaing,amr}.
Thus, identifying this latter equation
with Eq. (\ref{moment}), one finds  $\zeta(q) = F(-iq)$.

In the case of self-similar processes of exponent $H$, one easily gets that
the kernel is a dirac function
$G_{l,L}(u) =
\delta(u-H\ln(l/L))$ and consequently
$\zeta_q = qH$. The simplest non-linear (i.e., multifractal) case is the
so-called log-normal model that
corresponds to a parabolic $\zeta_q$ and a Gaussian kernel.  Let us note
that if $\zeta_q$ is non linear, a simple
concavity argument shows that Eq. (\ref{moment}) cannot hold for all $l$ in
$]0,+\infty[$.
Multiplicative cascading processes
\cite{man62,kp,ms,hent,wcasc}
consist in writing Eq. (\ref{castaingd}) starting
from some ``coarse'' scale $l = L$ and then
iterating it towards finer scales using an arbitrary fixed scale ratio
(e.g., $\lambda = 1/2$). Such processes
can be contructed rigorously using,
for instance, orthonormal wavelet bases \cite{wcasc}.
However, these processes have fundamental drawbacks: they do not lead
to 
stationary increments and they do not have continuous dilation invariance
properties. Indeed, they
involve a particular arbitrary scale ratio, i.e., Eq (\ref{moment}) holds
only for the discrete scales $l_n =
\lambda^n L$.

The goal of this paper is to build a multifractal process $X(t)$,
referred to as a Multifractal Random Walk (MRW),  with stationary increments
and such that Eq.
(\ref{moment}) holds for all
$l\le L$. 
We first build a discretized version $X_{\Delta t}(t=k\Delta t)$ of this
process. 
Let us note that the theoretical issue whether the limit process $X(t) =
\lim_{\Delta t \rightarrow
0} X_{\Delta t}(t)$ is well defined will be addressed
in a forthcoming paper. In this paper, we
 explain how it is built and prove that different quantities ($q$ order
moments, increment correlation,...) converge, when $\Delta t \rightarrow 0$.
 
Writing Eq.~ (\ref{castaingd})
at the smallest scale suggests that a good candidate might
be such that
$\delta _{\Delta t} X_{\Delta t} (k\Delta t) = \epsilon_{\Delta t}[k]
W_{\Delta t}[k]$ where $\epsilon_{\Delta
t}$ is a Gaussian variable and
$W_{\Delta t}[k] = e^{\omega_{\Delta t}[k]}$ is a log normal variable, i.e.,
\b
\label{X}
X_{\Delta t} (t) = \sum_{k=1}^{t/\Delta t} \delta_{\Delta t} X_{\Delta
t}(t) = \sum_{k=1}^{t/\Delta t} \epsilon_{\Delta t}[k] e^{\omega_{\Delta
t}[k]},
\e
with $X_{\Delta t}(0) = 0$ and $t=k\Delta t$.
Moreover, we will choose $\epsilon_{\Delta t}$ and $\omega_{\Delta t}$ to be
decorrelated and $\epsilon_{\Delta
t}$ to be a white noise of variance $\sigma^2 \Delta t$. Obviously, we need
to correlate the $\omega_{\Delta
t}[k]$'s otherwise $X_{\Delta t}$ would simply converge towards a Brownian
motion. 
Since, in the case of cascade-like processes, it has been shown
\cite{wcasc,ams,abmm} that the
covariance of the logarithm of the
increments decreases logarithmically, it seems natural to choose
\b
Cov(\omega_{\Delta t}[k_1],\omega_{\Delta t}[k_2]) =
\lambda^2 \ln \rho_{\Delta t} [|k_1-k_2|],
\e
with 
\b
\rho_{\Delta t} [k] =
\left\{
\begin{array}{ll} 
\frac{L}{(|k|+1)\Delta
t} & \mbox{for}~|k|\le L/\Delta t -1
\\
1 & \mbox{otherwise}
\end{array}
\right.,
\e
i.e., the $\omega_{\Delta t}$ are correlated up to a distance of $L$ and
their
variance $\lambda^2\ln(L/\Delta t)$
goes to $+\infty$ when $\Delta t$ goes to 0. For the variance of
$X_{\Delta t}$ to converge, a quick computation shows that we need to choose
\b
\label{r}
E\left(\omega_{\Delta t}[k]\right) = -rVar\left(\omega_{\Delta t}[k]\right)
= - r\lambda^2\ln(L/\Delta t),
\e
with $r=1$ (this value will be changed later) for which we find $Var(X(t)) =
\sigma^2
t$. 

Let us compute the moments of the increments of the MRW $X(t)$. Using the
definition of $X_{\Delta
t}(t)$ one gets
\begin{eqnarray}
\nonumber
\tiny
& E(X_{\Delta t}(t_1)...X_{\Delta t}(t_m)) =  \sum_{k_1=1}^{t_1/\Delta t}
...\sum_{k_m=1}^{t_m/\Delta t}  \\
\nonumber
\tiny
& E(\epsilon_{\Delta
t}[k_1]...\epsilon_{\Delta t}[k_m])E\left(e^{\omega_{\Delta
t}[k_1]+...+\omega_{\Delta t}[k_m]}\right).
\end{eqnarray}
Since $\epsilon_{\Delta t}$ is a 0 mean Gaussian process, this
expression is 0 if $m$ is odd. Let  $m=2p$.
Since the
$\epsilon_{\delta t}[k]$'s are
$\delta$-correlated Gaussian variables, one shows that the previous
expression reduces to
\[
 \frac{\sigma^{2p}}{2^p p!} \sum_{\s \in {S}_{2p}} \hskip -.3cm
\sum_{k_1=1}^{\scriptscriptstyle
(t_{\s(1)} \wedge t_{\s(2)})/\Delta t}
 \hskip -.3cm 
 .......
 \hskip -.3cm 
\sum_{k_{p}=1}^{\scriptscriptstyle (t_{\s(2p-1)} \wedge t_{\s(2p)})/\Delta
t}
\hskip -.8cm
E\left(e^{2\sum_{j=1}^{p}\omega_{\Delta
t}[k_j]}\right) \Delta t^p,
\]
where $a \wedge b$ refers to the minimum of $a$ and $b$ and
$S_{2p}$ to the set of the permutations on $\{1,...,2p\}$.
On the other hand, we have
$E\left(e^{2\sum_{j=1}^{p}\omega_{\Delta
t}[k_j]}\right)  = \prod_{i<j} \rho[k_i-k_j]^{4\lambda^2}.
$
Then, when $\Delta t \rightarrow 0$, the general expression of the moments
is
\begin{eqnarray}
\label{multicorr}
\nonumber
& E(X(t_1)...X(t_{2p})) = \frac{\sigma^{2p}}{2^p p!}
\sum_{\s \in {S}_{2p}} \int_{0}^{t_{\s(1)}\wedge t_{\s(2)}}du_1 \\
&  
...\int_{0}^{t_{\s(2p-1)}\wedge t_{\s(2p)}}du_p
\prod_{i<j} \rho(u_i-u_j)^{4\lambda^2},
\end{eqnarray}
where $\rho(t) = \lim_{\Delta t \rightarrow 0} \rho_{\Delta t} [t/\Delta
t]$. 
In the special case $t_1 = t_2 =... = t_{2p} = l$, a simple scaling argument
leads to the
continuous dilation invariance property
\begin{equation}
\label{modmom}
m(2p,l) = K_{2p} \left(\frac{l}{L}\right)^{p-2p(p-1)\lambda^2},~~\forall l
\le L,
\end{equation}
where we have denoted the prefactor
\[
K_{2p} = L^p \sigma^{2p} (2p-1)!! \int_{0}^1 du_1...\int_{0}^1 du_p
\prod_{i<j} |u_i-u_j|^{-4\lambda^2} \; .
\]
By analytical continuation, we thus obtain the following $\zeta_q$ spectrum
\begin{equation}
 \label{zetamodel}
\zeta_q = (q-q(q-2)\lambda^2)/2.
\end{equation}
\begin{figure}[h]
  \begin{center}
    \epsfig{file=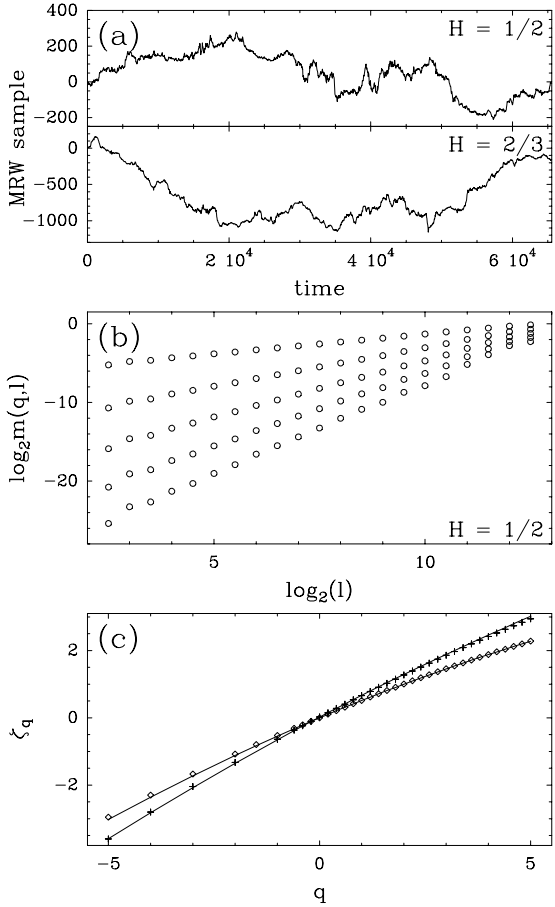,height=12cm}
    \caption{
      (a)  Plot of two realizations of $2^{17}$ samples of two MRWs with
$\lambda^2 = 0.03$, $L = 2048$ and where
$\epsilon_{\Delta t}$ is (top plot)
a white noise or (bottom plot) a fGn (Eq. (\ref{fgn})) with $H = 2/3$ .
      (b) Log-log plots of $m(q,l)$ of the MRW plotted in (a) (top plot)
versus $l$ for $q=1,2,3,4,5$.
      (c) ($\circ$) (resp. ($+$)) : $\zeta_q$ spectrum estimation of the MRW
plotted at the top (resp.
bottom) in (a). These estimations (obtained using the WTMM method
\cite{wtmm})
are in perfect agreement with the
theoretical predictions ($\ligne$) given by  Eq.~(\ref{zetamodel}) (resp.
Eq.~(\ref{zetamodel1})).
      }  
    \label{fig1}
  \end{center}
\end{figure}
We have illustrated this scaling behavior in fig. 1. Thus, the MRW $X(t)$ is
a multifractal process with
stationary increments and with continuous dilation invariance properties up
to the scale
$L$. Let us note that above this scale ($l >> L$), one gets from Eq.
(\ref{multicorr}) that
$\zeta_q = q/2$, i.e.,
the process scales like a simple Brownian motion, as if $\omega$ was not
correlated, though, of course, $X(t)$
is not Gaussian. Indeed,
 $K_{2p}$ is nothing
but the moment of order $2p$ of the random variable
$X(L)$ and is infinite for large $p$'s (depending on $\lambda$). Actually,
one can show that $\zeta_{2p} \le 0
\Rightarrow
K_{2p} = +\infty$. Consequently, the pdf of $X(L)$ has fat tails.
As illustrated in fig. 2, Eq. (\ref{castaing}) accounts very well
for the evolution of the pdf of the increments.
One shows that the
smaller the scale $l$, the fatter the tails of the pdf of $\delta_lX(t)$.
\begin{figure}[h]
  \begin{center}
    \includegraphics{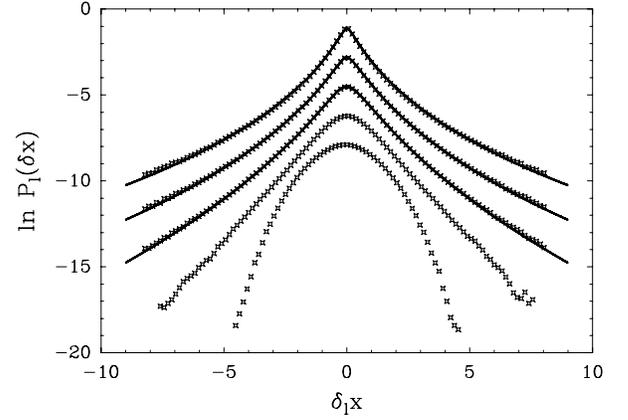}
    \caption{
      ($x$) Standardized estimated pdf's of $\ln \delta_l X(t)$
      for $l=4,32,256$, $2048$ and $4096$ (from top to bottom). These
estimations have been
made
       on 500 realizations of $2^{17}$ samples each of a MRW with $\lambda^2
= 0.06$ and 
      $L = 2048$. Plots have been arbitrarily shifted
      for illustration purpose.
       ($\ligne$) theoretical prediction from the estimated pdf at the
largest scale ($l = 2048$) using the
Castaing's equation (\ref{castaing}).
      }  
    \label{fig2}
  \end{center}
\end{figure}

Let us study the correlation structure of the increments of $X(t)$.
Since $\zeta_2 = 1$, one can prove that they are decorrelated
(though not independant).  Let
\begin{equation}
C_{2p}(l,\tau) =  <|\delta_\tau X(l)|^{2p} |\delta_\tau X(0)|^{2p}>,
\end{equation}
with $\tau < l$.
Using the same kind of arguments as above, one can show that
\begin{eqnarray}
\nonumber
& C_{2p}(l,\tau) = (\sigma^{2p}(2p-1)!!)^2 \int_l^{l+\tau} du_1 ...
\int_l^{l+\tau} du_p \\
& \int_0^{\tau} du_{p+1}...\int_0^{\tau} du_{2p} \prod_{1\le i<j\le2p}
\rho(u_i-u_j)^{4 \lambda^2}.
\end{eqnarray}
A straightforward argument then shows that
\[
K_{2p}^2 \frac{(\tau/L)^{2\zeta_{2p}}}{((l+\tau)/L)^{4\lambda^2p^2}} \le
C_{2p}(l, \tau) \le
K_{2p}^2 \frac{(\tau/L)^{2\zeta_{2p}}}{((l-\tau)/L)^{4\lambda^2p^2}},
\]
and consequently for $\tau << l$ fixed, using analytical continuation one
expects $C_q(l,\tau)$ to scale like
\b
\label{thecorr}
C_{q}(l,\tau) \sim 
K_{q}^2\left(\frac{\tau}{L}\right)^{2\zeta_{q}}\left(\frac{l}{L}\right)^{-\l
ambda^2q^2}.
\e
This behavior is illustrated in fig. 3.
 \begin{figure}[h]
  \begin{center}
    \includegraphics{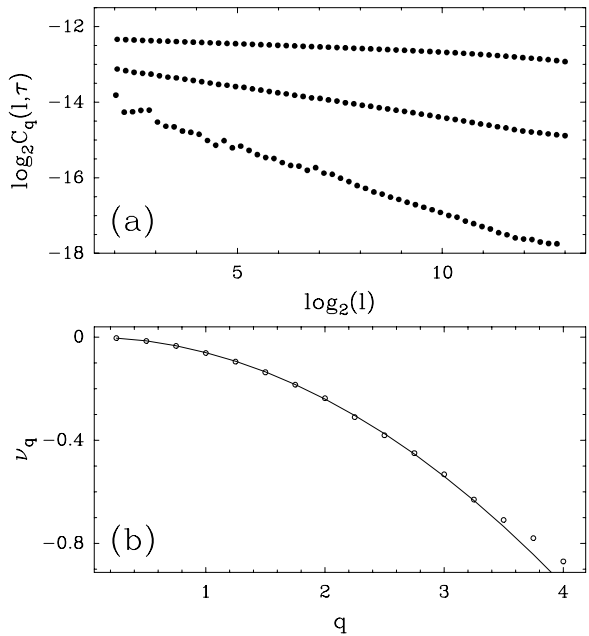}
    \caption{
      (a) Log-log plots of $C_q(l,\tau)$ versus $l$ for $q=1,2,3$.
      (b) Estimation ($\circ$) of the power law exponent $C_q(l,\tau)\sim
l^{\nu_q}$. It is in perfect
agreement with the theoretical prediction (Eq.
(\ref{thecorr})) $\nu_q = -\lambda^2 q^2$  ($\ligne$).
      }  
    \label{fig3}
  \end{center}
\end{figure}

>From the behavior of $C_q$ when $q \rightarrow 0$, we can obtain using
Eq. (\ref{thecorr}) that the covariance  of the logarithm of the increments
at scale $\tau$ and lag $l$
behaves  (for $\tau << l$) like
\begin{equation}
\label{logcorr}
C^{(\ln)}(l,\tau) 
 \sim -\lambda^2\ln\left(\frac{l}{L}\right).
 \end{equation} 
Thus, this correlation reflects the
correlation of the $\omega_{\Delta t}$
process and is the same as observed in Refs \cite{wcasc,ams,abmm}
for the cascade models. This behavior is checked in fig. 4.

\begin{figure}[t]
  \begin{center}
   \epsfig{file=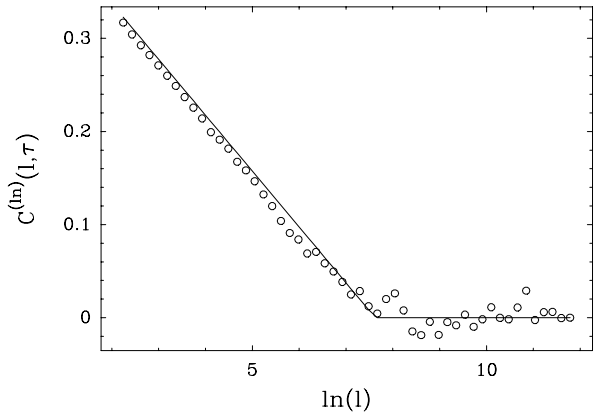}
    \caption{
      Estimation ($\circ$) of the correlation $C^{(\ln)}(l,\tau)$ of the
logarithm of the
increments. It is in perfect agreement with
the analytical expression (Eq. (\ref{logcorr})) $-\lambda^2
\ln\left(\frac{l}{L}\right)$
($\ligne$).
     }  
    \label{fig4}
  \end{center}

\end{figure}

Finally, let us note that,
one can built MRWs with correlated increments by just
replacing  the white noise
$\epsilon_{\Delta t}$ by a fractional Gaussian noise (fGn)
\b
\label{fgn}
\epsilon_{\Delta t}^{(H)}[k] = B_H((k+1)\Delta t)-B_H(k\Delta t),
\e
where $B_H(t)$ is a fBm with the scaling exponent $H$ and of variance
$\sigma^2t^{2H}$, and choosing $r = 1/2$ in
Eq.  (\ref{r}). Then, one can show (after tedious but straightforward
computations) that the spectrum of the MRW
$X^{(H)}(t)$ is
\begin{equation}
 \label{zetamodel1}
\zeta_q^{(H)} = qH-q(q-1)\lambda^2/2,
\end{equation} 
($\zeta_q^{(H)} = qH$ at  scales $>>L$) and consequently the MRW has
correlated increments.
Such a construction is illustrated in fig. 1 with $H=2/3$. Since $H>1/2$ it
leads to a process which is more regular
than the one previously built.

To summarize, we have built the MRWs,
a class of multifractal processes, with stationary increments and continuous
dilation invariance.
>From a theoretical point of view, MRW can
be seen as the simplest model that
contains the main ingredients for
multifractality, namely, logarithmic decaying correlation of
the logarithms of amplitude fluctuations.
Moreover, they involve very few parameters, mainly, the
correlation length $L$, the intermittency
parameter
$\lambda^2$, the variance
$\sigma^2$ and the roughness exponent $H$. They all can be easily estimated
using the $\zeta_q$ spectrum
and the increment correlation
structure.
We do believe that they should be very helpful in all the fields where
multiscaling is observed.
MRWs have already been proved successful for modelling financial data
\cite{fmuzy}. In this framework, we have shown that one can easily build
multivariate MRWs. Actually, the
construction of  MRWs,  as presented in this paper, can be used as a general
framework in which one can easily build
other classes of processes in order to match some specific experimental
situations. For instance,
a stationary MRW can be obtained by just adding a friction
$\gamma >0$,
 i.e.,
$X_{\Delta_t}[k] = (1-\gamma)X_{\Delta_t}[k-1] + \epsilon_{\Delta
t}[k]e^{\omega_{\Delta t}[k]}$. One can
build a strictly increasing MRW (and consequently a stochastic positive
multifractal measure) by just setting
$\epsilon_{\Delta t} = \Delta t$ in Eq. (\ref{X}) and use it as a
multifractal time for subordinating a monofractal
process (such as an fBm). One can also use other laws than the (log-)normal
for $\epsilon$ and/or $\omega$.
Another interesting point concerns the  problem
of the existence of a limit ($\Delta t\rightarrow 0$) stochastic process
and on the development of  a new stochastic calculus associated
to this process. All these prospects will be addressed in forthcoming
studies.

\newpage
We acknowledge Alain Arneodo for interesting
discussions. 

\vskip 2cm

\end{document}